# Predictive landscapes hidden beneath biological cellular automata

Lars Koopmans[1,2] and Hyun Youk[2-4]

[1]Program in Applied Physics, Delft University of Technology, Delft, The Netherlands
[2]Department of Systems Biology, University of Massachusetts Chan Medical School, Worcester, USA
[3]Program in Molecular Medicine, University of Massachusetts Chan Medical School, Worcester, USA
[4]Corresponding author: hyun.youk@umassmed.edu

**Abstract**
To celebrate Hans Frauenfelder's achievements, we examine energy(-like) "landscapes" for complex living systems. Energy landscapes summarize all possible dynamics of some physical systems. Energy(-like) landscapes can explain some biomolecular processes, including gene expression and, as Frauenfelder showed, protein folding. But energy-like landscapes and existing frameworks like statistical mechanics seem impractical for describing many living systems. Difficulties stem from living systems being high dimensional, nonlinear, and governed by many, tightly coupled constituents that are noisy. The predominant modeling approach is devising differential equations that are tailored to each living system. This ad hoc approach faces the notorious "parameter problem": models have numerous nonlinear, mathematical functions with unknown parameter values, even for describing just a few intracellular processes. One cannot measure many intracellular parameters or can only measure them as snapshots in time. Another modeling approach uses cellular automata to represent living systems as discrete dynamical systems with binary variables. Quantitative (Hamiltonian-based) rules can dictate cellular automata (e.g., Cellular Potts Model). But numerous biological features, in current practice, are qualitatively described rather than quantitatively (e.g., gene is (highly) expressed or not (highly) expressed). Cellular automata governed by verbal rules are useful representations for living systems and can mitigate the parameter problem. However, they can yield complex dynamics that are difficult to understand because the automata-governing rules are not quantitative and much of the existing mathematical tools and theorems apply to continuous but not discrete dynamical systems. Recent studies found ways to overcome this challenge. These studies either discovered or suggest an existence of predictive "landscapes" whose shapes are described by Lyapunov functions and yield "equations of motion" for a "pseudo-particle". The pseudo-particle represents the entire cellular lattice and moves on the landscape, thereby giving a low-dimensional representation of the cellular automata dynamics. We outline this promising modeling strategy.

---

## Introduction

Energy is both a fundamental concept and a practical tool. Often, students first encounter energy as a constant of integration, as a conserved quantity (total energy), that arises from integrating Newton's equation of motion. The integration also naturally defines the potential energy for a conservative force. The total and potential energies constrain a classical particle's past and future motion. Even when the particle dissipates energy, the potential energy landscape still lets us determine, at least in principle, the particle's past and future trajectories if we know the path-dependent dissipation function. Physics is a quantitatively predictive science largely due to the predictive powers of energy landscapes, conserved quantities, and generally applicable principles such as energy minimizations or entropy maximizations that guide the dynamics of physical systems. Coming up with similarly quantitative and predictive descriptions of living systems remains challenging. In this sense, living systems (e.g., cell, tissue) share similar challenges as complex non-living systems whose dynamics are also difficult to describe and predict without large-scale computer simulations (e.g., highly heterogenous (amorphous) materials; Earth's climate).

Living systems are difficult to model and when modeled, one often uses computer simulations or numerical solutions of nonlinear, coupled equations whose parameters cannot be measured or uniquely determined. Indeed, living systems are often high dimensional, nonlinear, and governed by many, tightly coupled constituents that are noisy. The predominant modeling approach, using differential equations that are tailored to each living system, thus faces the notorious "parameter problem": modeling just a few intracellular processes can involve numerous nonlinear, mathematical functions with unknown parameter values (e.g., three nonlinear, coupled equations with ten free parameters). One cannot measure many intracellular parameters or can only measure them as snapshots in time by chemically fixing the cells. As an example, even when a gene is regulated by just a handful of factors, we currently cannot analytically understand, without exhaustive numerical simulations, the precise gene-expression dynamics for many gene networks in a single cell (i.e., how many copies of mRNAs and proteins will be produced from each gene in a network as a function of time). The difficulty arises from the stochastic and deterministic equations that describe gene-expression dynamics quickly becoming analytically intractable as one increases the number of genes or transcriptional regulators in the gene network. For example, a realistic network of just four genes, with each one controlled by three or more transcription factors, can be analytically intractable. Even for a handful of genes, there can be many mathematical functions as candidates, many of which are nonlinear [1]. One can simulate the dynamics of gene networks and other cellular processes (e.g., metabolic fluxes). But even so, the numerical simulations are often based on myriad parameters whose values cannot be uniquely inferred or measured in a living cell [2]. Consequently, one often creates, ad hoc, a numerical simulation that is tailored for a particular cellular process and then evaluates how sensitive the simulation is to large changes in the parameter values (i.e., "parameter-sensitivity analysis"). The situation that we now have is that each simulation of a cellular process can seem disjointed from the others. It is therefore of interest to look for other approaches to modeling living systems and, more importantly, hidden principles that apply to multiple, seemingly unrelated, living systems. Such approaches may help us get around the ad hoc, brute-force simulations that are typical in modeling living systems today.

Traditional quantities of physics such as position, momentum, and energy are neither measurable nor the most sensible quantities to describe networks of genes and cellular processes. A more appropriate view of living systems and their chemical subsystems, such as gene networks, is that they are complex dynamical systems described by a set of reaction-rate (differential) equations. This is like viewing a system of particles as a dynamical system. Viewing particle mechanics as a dynamical system, after the invention of Newtonian mechanics, led to the Hamilton's equation of motion. Hamilton's equation of motion recapitulated the Newtonian mechanics and showed a deeper, mathematical (geometric) structure that underlies classical mechanics. Similarly, by viewing living systems as dynamical systems, we can ask whether there exists a mathematical structure for dynamical systems that is generalizable to a large class of cellular processes. There has been notable progress in this direction, namely progress on chemical networks. Notable examples include the theory of dissipative systems by Gregoire Nicolis, Ilya Prigogine, and their co-workers [3]. Another notable example of progress on (bio)chemical systems has been made by mathematicians such as Martin Feinberg, Friedrich Horn, and Roy Jackson who have built and advanced the chemical reaction network theory [4]. The chemical reaction network theory makes general statements, such as the number and stability of equilibrium (fixed) points of a dynamical system that represents a network of chemical reactions. The theory does so with a graph representation of chemical reactions in which a node represents a chemical specie while a directed edge represents a chemical conversion. This theory also shows that functions like energy, a function that is non-increasing over time, exist for a set of chemical reactions whose graph structure satisfies certain criteria. This result is called the "deficiency zero theorem" [4]. Existing theories like the chemical reaction network theory provide deep insights into chemical reactions. Although they can help us understand well-defined chemical reactions in test tubes, applying them to cellular processes has been challenging because: (1) we often do not know the graph structure (connectivity) between the various molecular species inside a cell; (2) we know that the connections of such a graph would change over time in a cell but we do not know how; and (3) we do not know how to relate a phenotype to a genotypic information such as the steady-state expression levels of genes and their stability. Perhaps this last point is one of the most difficult challenges for physics and biology: how to turn numbers into life.

In this perspective paper, we explore the idea that living systems are dynamical systems and whether predictive landscapes like energy landscapes exist for describing these dynamical systems. We will begin by reviewing, for both biologists and physicists, how energy landscapes describe non-living (physical) systems. We will then describe recent studies that viewed multicellular systems as dynamical systems and used cellular automata to model how cells regulate each other's gene expression [5-7]. These studies either discovered or suggest an existence of a predictive, energy-like landscape that produces an "equation of motion" for spatial-pattern-forming dynamics in a field of finite numbers of cells as well as other predictive, statistical metrics for the cellular automaton dynamics. We will focus on cellular automata that follow verbal (qualitative) rules in which cells have discrete states rather than a continuum of states. Our focus on such cellular automata is due to a practical reason: much of the literature in biology use qualitative descriptions of cellular processes. That is, much of the literature in biology use qualitative statements to describe living processes that one wishes to model. For example, a transcription factor either binds (strongly binds) or does not bind (weakly binds) to a particular location on a chromosome. A gene is either expressed (highly expressed) or not expressed (lowly expressed) [8,9]. Such binary statements capture some of the most salient features of living systems in terms of discrete variables (i.e., variables with two or finitely more states). These binary statements are approximations of the real phenomena. Cellular automata easily capture these binary statements. The rules of cellular automata can be qualitative statements that do not require any knowledge of rate constants, thus seemingly avoiding the "parameters" problem that plagues modeling of living systems with differential equations. Several other approaches, without involving cellular automata, also mitigate the parameters problem while still using a system of nonlinear equations [10,11]. But qualitative rules of cellular automata have the advantage of easily describing living systems, much of which are still predominantly studied qualitatively rather than quantitatively. The challenge then is to quantitatively understand biologically motivated cellular automata whose rules are written in verbal, binary statements. It is challenging to predict dynamics of such cellular automata. For one, much of today's mathematical machineries are designed for continuous dynamical systems (e.g., complex analysis, differential equations, chemical reaction network theory). Relatively few mathematical tools and general theorems exist for analyzing cellular automata [12]. We will describe recent studies that show ways to overcome these challenges. We will end this paper by suggesting ways of advancing our understanding of cellular automata that model spatial patterning dynamics by cells that communicate with each other through diffusible molecules.

**Predictive landscapes for a particle and a gene**

Potential energy landscapes in classical physics are useful because they simplify our analysis of a particle's motion by turning a vector problem into a scalar problem. Specifically, a potential energy landscape produces a vector field that dictates the force on a particle at every location. The vector field visually exemplifies the predictive power of physics: if we know the particle's current velocity and position, then we can determine both its past and future trajectories. Analogously, one can also build predictive vector fields that describe gene-expression dynamics but only for sufficiently simple gene networks. Here, a vector field represents changes in the amount of mRNA or proteins for each gene, with the dimensionality of the space ("gene-expression space") being equal to the number of genes in the network. A network of just four genes can be beyond our ability to visualize. For a network of three genes, the 3D space can be visualized but the underlying vector field may be difficult to determine due to the nonlinearity of the equations that govern the gene regulatory network and the parameters that those equations use, whose values may be difficult to determine.

A single scalar function (potential-energy function) produces vector fields that one can describe with differential equations. However, dynamical systems that represent gene expression do not lead to potential-energy-type functions because they lack the time-reversal symmetry. Hence, there may be no reduction of a vector problem into a simpler scalar problem for an arbitrary gene network. Aside from the direction of motion, potential energy landscapes reveal all possible equilibria of a particle and their stability. Similarly, vector fields that describe gene-expression dynamics also reveal equilibrium levels of gene expression (i.e., steady-state levels of mRNAs or proteins). These may be stable or unstable equilibria - they are fixed points of a dynamical system - just like the equilibrium positions of a physical particle. Interestingly, one often finds more "exotic" fixed points for gene expression. For example, a fixed point of a gene network dynamics can exhibit excitability whereby a small perturbation in expression level leads to an excursion far away from the fixed point in the phase space that lasts for a long time - the initial perturbation does not exponentially decay over time - before the system (gene expression levels) returns to the fixed point [13,14]. Similar phenomena also arise in physical oscillators and laser dynamics. One encounters such cases for gene expression dynamics because of the numerous, nonlinear functions that govern the production and degradation of biomolecules. This highlights one of the main challenges of designing a common (not ad hoc) algorithm for modeling living systems: instead of a generic function, we need to consider the specific rates and numerous mathematical functions even for a few biomolecular processes. Moreover, these mathematical functions are typically phenomenological in that they describe "lumped" actions of many players such as RNA polymerases and ribosomes. It is difficult to relate these phenomenological equations to established, traditional quantities from physics such as energy, entropy, and temperature. This situation presents another challenge for biological physicists who look for generalizable principles that underlie living systems [15].

To describe the stochastic production and degradation of RNAs and proteins, one typically uses a Master equation. But Master equations are difficult to solve. To make progress, one approximates the Master equation with a Fokker-Planck equation. The steady-state solution to a Fokker-Plank equation can yield an energy-like function [16-18]. But only in very special cases can one express the steady-state solutions of Fokker-Planck equation in terms of a conservative flux. Concretely, in the case of a single gene that regulates its own expression with an autoregulatory positive feedback loop, the steady-state distribution for the copy number of mRNAs is determined by an "energy landscape" that resembles a double-well potential, with each of the two wells corresponding to two stable, steady-state expression levels. Ultimately, however, such a landscape from Fokker-Planck is not generalizable. For one, the landscape only works for steady-state gene expression and one does not need a landscape formulation to determine the steady-state expression level. Finding the landscape also requires solving the Fokker-Planck equation at a steady state, which is not analytically tractable in many situations. Recently, researchers have proposed some promising approaches for deducing similar landscapes without needing to solve the Fokker-Planck equation. One of them involves sampling enough stochastic trajectories in the gene-expression space - generated by iteratively running Monte Carlo simulations of the Master equation for the gene network - and then deducing the main features of an "energy landscape" that would produce these trajectories [19]. Given that Fokker-Planck dynamics in high dimensions is generally not computationally tractable, this is a promising direction for deducing landscapes that produce predictive vector fields in a gene-expression space.

Fundamentally, however, it is not obvious that living systems conserve some quantity, essentially because they are out of thermal equilibrium. A deeper reason for conservation of energy arising as a constant from integrating Newton's equation is that the system's Lagrangian is invariant under translation in time. This is just one application of Noether's theorem, which states that a particular symmetry in a system, such as time-translation invariance or rotational invariance, necessarily yields a conserved quantity such as energy (for time-translation invariance) and angular momentum (for rotational invariance). It is unclear that there is any symmetry in biological processes such as gene expression or metabolism. Hence, if there are conserved quantities that we could exploit to predict cellular dynamics, then such quantities do not arise from any (obvious) symmetry in the system. In fact, living systems are disordered systems with a clear (forward) direction in time. Hence it is difficult to define any helpful symmetry to formulate a conserved quantity. Between the two challenges - the lack of any obvious symmetry and being out of thermal equilibrium - researchers have made progress on the latter by exploiting the idea of quasi steady states. We turn to this in the next section.

**Landscapes for particles and genes with slow changes**
When describing macroscopic systems with many particles, specifying the position and velocity of each particle is infeasible. Instead, in statistical physics, we can describe the collective behavior of particles in a probabilistic sense. For example, if we have a box of ideal gas, we can define a phase space, representing the position and moment of each particle, that captures every degree of freedom for the system. Then, in thermal equilibrium, we can show from first principles that the particles' velocities are distributed according to the Boltzmann distribution. However, if a system is not in thermal equilibrium, the total energy changes over time and we would need a different approach to analyze the system. One such approach uses the potential energy landscape. For example, Potential Energy Landscape (PEL) has been successfully used to address a major challenge in condensed matter physics: explaining amorphous solids such as glass-forming materials [20-24]. Next to the complexity that the irregular positions of the atoms bring, the slow liquid-like motion of the atoms on larger time scales leads to a non-equilibrium system. For such materials, one can define a PEL in terms of the position of the atoms and their interactions. This leads to a rugged, complex, high dimensional energy landscape [21-24]. This landscape can then be drastically simplified by only describing its geometric, global properties such as any local minima that it has, the distance and barrier heights between the local minima, and the landscape's curvature. Such a reduction in information allows one to successfully model glass-forming systems and explain their macroscopic properties [20-24].

A more biologically relevant example of using energy landscapes to understand out-of-equilibrium dynamics is in protein folding [25-30]. After translated by ribosomes, the resulting polypeptide folds into a specific three-dimensional shape to properly function in a cell. Reminiscent of the PEL approach, one can define a free energy landscape for this folding process. The landscape, like PEL for glass-forming systems, is also a rugged, funnel-shaped landscape with the native conformation corresponding to the bottom of the energy landscape. One can then apply this energy landscape to consider a protein to be adiabatically folding and thereby apply (quasi steady state) equilibrium thermodynamics to describe the folding dynamics. This landscape view helps us understand the protein-folding dynamics and in characterizing all possible structures that a given sequence of peptides can form. Hans Frauenfelder, whom our article honors as part of the journal's special issue, is one of the pioneers of this approach to understanding protein-folding dynamics [26-30].

Despite the success of statistical-physics approaches for studying protein-folding and molecular dynamics, it has been difficult to develop a similar approach for explaining living systems whose behaviors are not obviously described in terms of physics quantities such as position, velocity, energy, and the likes. Gene-expression dynamics, whose more natural description is in terms of phenomenological variables such as the rates of production, degradation, binding, and unbinding, is just one example that remains challenging to model without resorting to exhaustive numerical simulations in many cases. Researchers have made some progress by using equilibrium statistical mechanics in which one uses the binding energies of transcription factors to determine steady-state gene-expression levels in microbes [31-38]. The main reason for the success of this approach is that certain processes, such as the binding and unbinding of a transcription factor at a target locus on DNA, occur much faster than the other processes involved in gene expression. Hence, one can assume that transcription factors obey an equilibrium (Boltzmann) distribution as a function of their binding energies. The challenge here is that one usually cannot obtain the binding energies of transcription factors inside a cell, even for one gene. Moreover, not all gene-expression dynamics have clearly separable timescales. Nonetheless, using equilibrium statistical mechanics with a quasi-steady-state approximation is a promising approach to studying gene-expression dynamics.

**Predictive landscapes for living systems beyond gene networks**
For cellular processes beyond gene expression dynamics in a cell, we do not yet know how and whether the concept of energy or an energy-like function can play a useful role. In principle, one can define the free energy of individual molecules that are involved in gene expression (e.g., RNA polymerases). But it is unclear how knowing the free energy can simplify the analysis. When searching for energy or some other function like it, we're really talking about Lyapunov functions. Lyapunov functions are mathematical functions that monotonically change (non-increasing or non-decreasing) over time. There is no general algorithm for constructing a Lyapunov function for a dynamical system or for determining whether one can even exist for a given system [39]. Thus, it is nontrivial to find a Lyapunov function, if it exists, for a living system. Moreover, we do not yet know of a general algorithm for determining how a cell will behave over time. This is unsurprising since a cell can do many things and the mathematical function would need to be tailored for the cellular feature that one wishes to study. If we examine the examples of energy landscapes for non-living systems, we see that potential energy is defined in terms of either physical location of a particle or, more generally, the coordinates that we can assign to each degree of freedom in a system. In a cell, it is generally unclear what the "degrees of freedom" are, how we would measure these, and how these would be mathematically represented. In fact, it is not obvious that we can use metrics from physics such as pressure, energy, and position to describe some of the most important cellular and organismal behaviors, many of which do not originate from mechanical means. The combination of so many mechanical and chemical parts that lead to a phenotype can also blur the contribution of each, even those that involve mechanical aspects of a cell that physicists are more familiar with explaining. Although there may not be a conserved quantity for most cellular processes, there may still be nontrivial, mathematical functions that can yield a predictive, "equation of motion" for the cellular process of interest. In our view, this is ultimately what a biological physicist can realistically hope for. Recent studies by our group have found such functions for cellular automata that self-organize spatial patterns. We describe these in the next section.

**Predictive landscapes for cellular automata that self-organize spatial patterns**
Continuous dynamical systems are described by differential equations. They are biologically relevant, interesting, and pose many challenges. A less well explored approach is modeling living systems as discrete dynamical systems. There are also less mathematical tools for analyzing discrete dynamical systems, such as those modeled by cellular automata, than there are for analyzing continuous dynamical systems [12]. For example, the whole of mathematical (complex) analysis is more naturally suited for analyzing continuous dynamical systems than discrete dynamical systems. Cellular automata are natural choices for modeling discrete dynamics of multicellular systems as they use individual agents that each have discrete states and interact with their nearest-neighboring cells. [40,41].

Discrete systems are suitable for modeling living systems whose dynamics involve two or more processes that occur with differing timescales (i.e., fast and slow). This means that we can assume that the faster processes equilibrate first and then the slower processes build on top of (or respond to) the already-equilibrated, faster processes. As the slower processes change, however, they can perturb the quasi-steady states reached by the faster processes, causing the faster processes to exit their equilibria and then equilibrate in new states. This leads to a back-and-forth between the fast and slow processes, causing the entire system to stay out of equilibrium for a long time, potentially forever. Such back-and-forth dynamics are common for living systems. These bidirectional feedbacks are often nonlinear in nature, making their analyses difficult. Multicellular systems frequently exhibit these back-and-forth behaviors [42-45]. For example, cells can communicate with one another by secreting molecules that diffuse from one cell to another. The molecules can diffuse on a timescale that is shorter than the timescale at which gene expression changes occurs due to a cell sensing the molecules. Hence, we can consider the secreted molecules to reach a steady-state concentration and the cells reacting to it by, say, turning on or off gene expression after sensing the molecules. One of these genes may encode the secreted molecule, meaning that the cells can, in turn, change the rate at which they secrete the molecules or change the expression level of the receptor that senses the molecule. Such "secrete-and-sense cells" are ubiquitous in nature [42,44,45]. A cellular automaton is ideal for modeling such systems.

A cellular automaton can be considered as a computer. It evolves over time by applying a pre-defined algorithm to the initial configuration that we assign to the "cells" on a grid. We can consider the initial configuration of the cellular lattice to be an "input data" for a computer program [40]. Cellular automata reveal how iteratively applying the same rules can lead to complex dynamics. This is a deep notion in computational theories and biology. Perhaps one of the most well-known cellular automata is John Conway's Game of Life [46]. The Game of Life has captured the attention of the public and scientists alike because it is Turing complete, meaning that anything that a computer can do can also be programmed into the initial configuration assigned to the Game of Life [47]. Loosely, Turing completeness means that anything can happen in the Game of Life if the user can determine the initial configuration that will produce the desired dynamics. For modeling multicellular systems, one of the most widely used cellular automata is the Cellular Potts Model (CPM) [48,49]. In CPM, adjacent cells mechanically interact by exerting force on one another, through the cell-cell contacts. A Hamiltonian captures the cell-cell interaction energies in a simplified and insightful way. The CPM then evolves over time with the rule that the Hamiltonian must be non-increasing over time, leading to a final configuration of the tissue in which the Hamiltonian takes on a locally minimal value. CPM is an example of a cellular automaton in which the Hamiltonian is built in as the rule.

Unlike CPM, many cellular automata do not have a built-in Hamiltonian that dictates the dynamics. Instead, verbal (qualitative) rules dictate their temporal evolution as in the Game of Life. It is a priori unclear whether a predictive, energy-like landscape or any Lyapunov function exists for such cellular automata and if so, how one might find them [39]. In a recent work, our group showed that such a landscape exists for a cellular automaton that models spatial-patterning dynamics of a field of secrete-and-sense cells [6] (Fig. 1A). In this "secrete-and-sense automaton", cells communicate by secreting and sensing one type of diffusing molecule called a "signaling molecule" [5,6]. A cell can secrete multiple copies of the signaling molecule and sense the concentration of the molecules on itself. The concentration of the signaling molecule on a cell determines whether that cell turns on the expression of a gene (if the concentration is above a certain threshold value) or turns off the expression (if the concentration is below the threshold value). Thus, a cell can be in one of two states: "ON" or "OFF". With $N$ cells, the cellular automaton can start with $2^N$ possible states. When a cell is ON, then it secretes the signaling molecule at a "high" rate whereas it secretes the signaling molecule at a "low" (basal) rate if the cell is OFF. Our group found that this cellular automaton always terminates with a configuration (spatial configuration of ON and OFF cells) that is more spatially ordered than the initial configuration [5,6] (Fig. 1A - left box). Although a reaction-diffusion equation governs the steady-state concentration of the signaling molecule on each cell, the temporal evolution of the cell states can be mapped onto verbal rules. This is because of the binary nature (high/low) of the secretion rate for the signaling molecule and the fixed distances between the immotile cells. We can convert the reaction-diffusion equations into verbal rules that dictate the cellular automaton such as, for example, "a cell adopts an ON-state if at least four of its nearest neighbor cells are in the ON-state". Changing two control parameters of the system - the threshold concentration *K* and the maximal secretion rate $S_{ON}$ of the signaling molecule – alters these verbal rules (e.g., turning on a cell now requires three nearest neighbors as ON-cells, not four). In this cellular automaton, one can use a phase diagram that is determined as a function of *K* and $S_{ON}$ to summarize how these verbal rules that run the automaton change (e.g., a cell turns ON if there are at least four nearest-neighbor cells that are ON) [5,6].

In the secrete-and-sense cellular automaton, we quantified the spatial order - how ordered the multicellular configuration of ON and OFF cells is - by defining a weighted spatial autocorrelation function called the "spatial index" *I* [5,6]. The spatial patterns generated by the cellular automaton include stripes and islands formed by ON-cells clustered together, as seen in real secrete-and-sense cells in nature. Moreover, two "macrostate variables" simplify the description of the spatial-pattern-forming dynamics of the cellular automaton: the fraction *p* of cells that have the gene of interest in the ON state and the spatial index *I*. With these two macrostate variables, one no longer needs to keep track of which of the hundreds of cells in the lattice are ON and which are OFF. A trade-off, however, is that one now must resort to a probabilistic description of the cellular automaton dynamics since each macrostate (*p, I*) can have numerous (thousands or more) microstates (i.e., one of the $2^N$ spatial configurations). Crucially, one can find a predictive landscape that approximately describes how the cellular automaton evolves over time in the two-dimensional space spanned by *p* and *I* [6] (Fig. 1A - right box). One can visualize this "pseudo-energy landscape" because it is a function only of the two macrostate variables: *p,* and *I.* The landscape, at once, reveals all possible final configurations and the probability with which the cellular automaton terminates at each configuration. The pseudo-energy landscape also reveals, to a good approximation, how the multicellular configuration evolves over time. One can write the pseudo-energy function in terms of the microstate variables (i.e., the ON/OFF state of every cell). It then mathematically resembles a Hamiltonian for a spin system with non-nearest neighbor interactions. In the cellular automaton, the strength of the interaction between two communicating cells is defined by a "signal strength" function that is a function of the distance between the two cells (spins) and other parameters. One can rigorously prove that this function does not increase over time, thereby proving that the function is a Lyapunov function like physical energy [6]. But unlike a physical energy, the pseudo-energy function does not provide a definite, stopping condition for the cellular automaton. Instead, there is a "trapping probability" assigned to each location on a landscape: the probability that the cellular automaton terminates for each value of (*p, I*). Intuitively, this means that the pseudo-energy landscape is adhesive, with the adhesiveness (trapping probability) shown as a heatmap color painted on the landscape (Fig. 1A - right box). The shape (gradient) of the landscape represents the direction in which the particle, which represents the configuration of all cells - rolls down. This represents the temporal evolution of the macrostates of the cellular automaton. These features make the pseudo-energy landscape useful to understand the dynamics of the secrete-and-sense cellular automaton. It provides a probabilistic view of the dynamics because the macrostates - *p* and *I* - do not define the stopping condition of the cellular automaton. Moreover, many microstates can belong to one macrostate (*p, I*). The reduction in dimensions from $2^N$ to two (i.e., *p* and *I*) comes at the expense of relinquishing determinism for a probabilistic description. Finally, one can precisely determine how changing the verbal rules of the cellular automaton changes the pseudo-energy landscape's shape and the adhesiveness at each location of the landscape [6].

In terms of one of the macrostate variables, the fraction $p$ of cells that are ON, and given the location in the phase diagram (i.e., the values of $K$ and $S_{ON}$), a mean-field approach lets one derive an analytical formula for the total number of spatial configurations that a field of cells can have (i.e., the total number of terminal configurations of the cellular automaton). We first derived this formula, which we call "multicellular entropy", for the secrete-and-sense automaton with one type of molecule (i.e., with one cell type and one type of secreted molecule [5]). We later generalized the multicellular entropy to secrete-and-sense automata involving any number of cell types and any number of secreted molecules (e.g., three types of cells with two molecules whereby each cell type has a different $K$ and $S_{ON}$ for each molecule type) [6]. For the simplest secrete-and-sense automaton, which has just one type of secreted molecule and one type of cell, all initial configurations lead to the cellular automaton eventually halting with a more spatially ordered configuration (e.g., clustered island of ON-cells).

One primarily uses energy landscapes to determine a system's equilibrium states. Turing showed that there cannot be a general method that determines whether an arbitrary cellular automaton (computer program) will definitively halt [50]. This is known as the halting problem. It is thus likely that there cannot be a general method that takes an arbitrary cellular automaton as an input and then constructs a mathematical function, like an energy landscape, that definitively predicts the configurations with which the cellular automaton will terminate. At best, we can hope to give the probability of the automaton stopping for each configuration. The adhesive pseudo-energy landscape does this for the secrete-and-sense automaton with one type of signaling molecule.

The cellular automaton for secrete-and-sense cells that communicate with two distinct signaling molecules display dynamics that markedly differ from those of the one-molecule cellular automaton. The most interesting dynamics occur when the two molecules regulate each other's secretion. Most notably, the cellular automaton with two molecules may not terminate at all or not terminate within a reasonable, observable time on a computer whereas the cellular automaton with one molecule always terminates for any initial configuration. In particular, the two-molecule cellular automaton with just 100 to 200 cells can lead to a perpetually traveling wave (Fig. 1B - left box) or puddles of spiral waves that constantly changes over time until they start to periodically cycle through a set of states indefinitely. These "dynamic spatial patterns" self-organize from an initially disordered field of cells. Although a "landscape" has not yet been found for these types of cellular automata, we found hints for its existence [7] (Fig. 1B). Specifically, the start of any process by which the dynamic spatial patterns form seems to resemble a ball, representing a cellular lattice configuration, rapidly rolling down a side of a bowl and landing on the bowl's a flat bottom (Fig. 1B - right box). Height of a bowl represents the amount of spatial order (quantified by the spatial index *I*). A decreasing height represents an increasing spatial order. Hence the ball rapidly rolling down the side of a bowl represents the cellular lattice quickly becoming more spatially ordered at the start of a process that generates a dynamic spatial pattern. After this stage of the process, the lattice configuration appears to erratically and randomly fluctuate without becoming either more or less spatially ordered. We can think of these fluctuating dynamics to resemble a billiard ball bouncing around, erratically and randomly, on the bowl's flat bottom. Eventually and seemingly without any warning, the lattice configuration undergoes a rapid becomes more spatially ordered and forms a traveling wave or some other highly organized, dynamical spatial pattern. This is as if the particle suddenly falls through a small crack somewhere on the flat bottom after randomly bouncing around for a while (Fig. 1B - right box).

While this picture is not quantitative and is merely an analogy, we found several features in the two-molecule cellular automata that strongly suggest some landscape like the bowl may underlie the two-molecule cellular automata that generate dynamic spatial patterns [7].

**Outlook**

Cellular automata are useful in modeling discrete dynamics of living systems. As an example of its relevance, experimentalists have recently discovered that the lizard skin uses a hexagonal lattice of cells (like the triangular lattice of cells in [5-7]) to pattern the skin color and that this real-life patterning dynamics is a cellular automaton [51]. Use of cellular automata in understanding living systems has a long history, starting with Von Neumann's cellular automaton that yields self-replicating structures [52]. Commonly used cellular automata for modeling biological dynamics usually have built-in Hamiltonians that dictate how the cells in the automata change their states over time. Minimizing this Hamiltonian over time is imposed as a rule for the cellular automaton. This is the case in Cellular Potts Model [48,49]. Moreover, physicists have typically used cellular automata to model mechanical aspects of living systems (e.g., Vicsek model for motion of birds [53] and the CPM for mechanical shuffling of cells in a tissue [48,49]). A relatively underexplored avenue of research is using cellular automata that do not have any built-in Hamiltonians and are not describable by mechanical, electrical, or any other traditional quantities of physics (e.g., position, velocity, forces, energy). An example of this is the cellular automata for secrete-and-sense cells that regulate each other's gene expression by secreting and sensing signaling molecules [5-7]. We believe that insights into biology can arise from studying this latter class of cellular automata which are governed by "simple" verbal rules. One of the major advantages of cellular automata with qualitative statements as rules that govern local cell-cell interactions is that much of the salient biological features are verbally described as discreet (binary) approximations (e.g., transcription factor either (strongly) binds or does not (strongly) bind to a promoter) [7, 8]. Our viewpoint here is like that of others, including those who have recently described the advantages of discretizing (binarizing) features of biomolecular networks to build models that reveal the essential features of living systems [54-57]. Qualitative rules of cellular automata also do not require knowing the values of myriad parameters, many of which are difficult to ascertain in living systems. Cellular automata based on qualitative rules that do not start out with a built-in Hamiltonian may unveil new connections between theories of computational complexity and biology. Some outstanding questions are: What are the classes of dynamics that can arise for each type of biologically motivated rule in a cellular automaton? How do the cellular automaton dynamics change if we mutate each rule? Can objects such as pseudo-energy landscapes underlie cellular automaton dynamics as is the case for the secrete-and-sense cells? We expect that addressing these questions for biologically motivated cellular automata will reveal a comprehensive classification of all possible cellular dynamics that can arise from local cell-cell interactions. Ultimately, we suspect that this more abstract line of investigation may reveal deep links between living systems and theories of computation.

**Acknowledgements**

We thank Robert Brewster for helpful comments. H.Y. was supported by the Netherlands Organisation for Scientific Research (NWO) Vidi award (no. 680-47-544), CIFAR Catalyst Grant, and EMBO Young Investigator Award.

**Compliance with Ethical Standards:**

The authors declare that they have complied with all ethical standards regarding this paper.

**Competing interests**

The authors declare no competing interests.

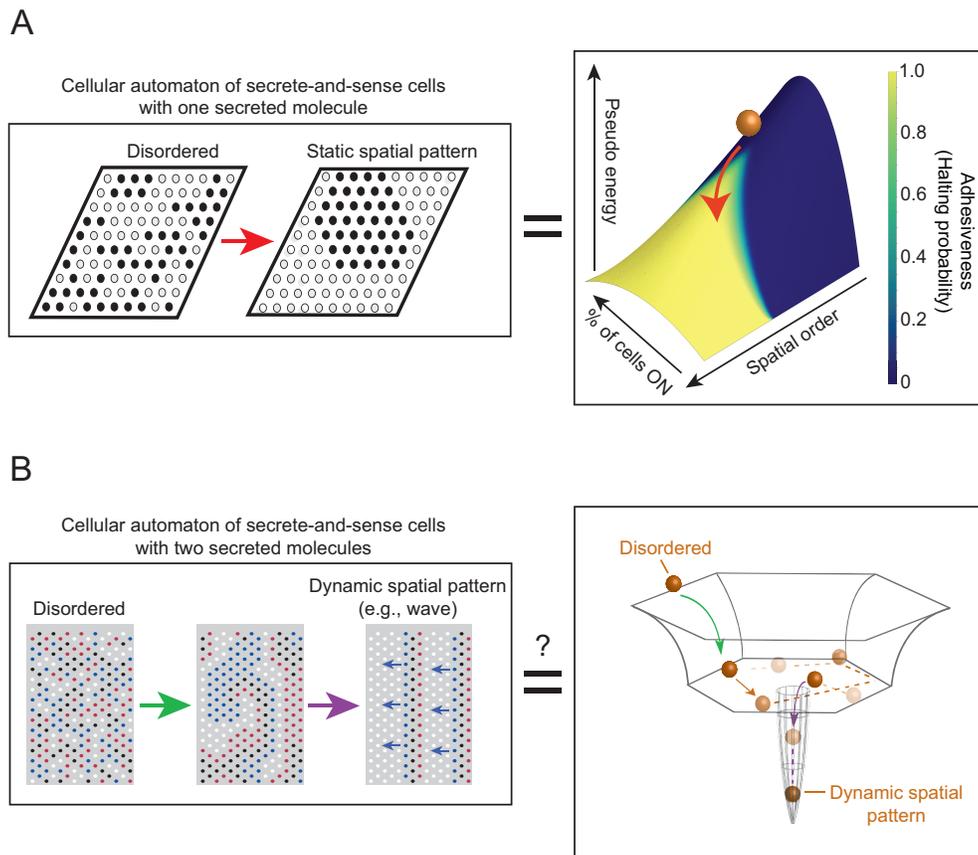

**Fig. 1. Predictive landscapes for cellular automata that self-organize spatial patterns via cell-cell communication.**
**(A)** Left box: Cellular automaton in which each cell (circle) has two possible states (2 colors) and secretes one type of signaling molecule that activates its own secretion [6]. This cellular automaton always terminates with a static configuration that is more ordered than the initial configuration. Right box: Pseudo-energy landscape in which a ball represents the entire cellular lattice (shown in left box). The ball rolling down the landscape represents the temporal evolution of cellular automaton. The landscape is sticky, causing the ball to stick with some probability at each location (shown in color bar) which represents cellular automaton halting.
**(B)** Left box: Cellular automaton in which each cell (circle) has four possible states (4 colors) and secretes two types of signaling molecules, each of which either activates or represses the secretion of itself or the other molecule [7]. This cellular automaton does not always terminate because it can form a never-ending (looping) dynamic spatial pattern such as a perpetually traveling wave (rightmost configuration in left box). Right box: Hypothesized predictive landscape for this cellular automaton.